\documentclass[11pt,tightenlines,eqsecnum,floats,aps,nofootinbib,prd]{revtex4-2}
\usepackage{hyperref}
\usepackage{amsmath,amssymb,amsfonts,amsthm,amscd}
\usepackage{enumerate}
\usepackage{colordvi}
\usepackage{units}
\usepackage{epsfig}
\usepackage{natbib}
\usepackage{enumerate}
\usepackage{afterpage}
\usepackage{physics}
\usepackage[utf8]{inputenc}
\usepackage[parfill]{parskip}
\usepackage{tikz}
\usetikzlibrary{positioning}
\usepackage{caption}

\def\be{\begin{equation}}
\def\ee{\end{equation}}
\def\ba{\begin{eqnarray}}
\def\ea{\end{eqnarray}}

\def\Lie{\mathfrak{L}}

\def\D{\mathbf{D}}

\def\d{\mathrm{d}}

\begin{document}

\title{Dynamical Similarity in Field Theories}

\author{David Sloan}
\email{d.sloan@lancaster.ac.uk}
\affiliation{Lancaster University}
\begin{abstract}
In previous work I have shown that Herglotz actions reproduce the dynamics of classical mechanical theories which exhibit dynamical similarities. Recent work has shown how to extend field theories in both the Lagrangian and de Donder-Weyl formalism to contact geometry \cite{gaset2020contact,gaset2021k,gaset2022herglotz}. In this article I show how dynamical similarity applies in field theory. This is applied in both the Lagrangian and Hamiltonian frameworks, producing the contact equivalents. The result can be applied to general relativity where I demonstrate how to construct a complete description of the dynamics, equivalent to those derived from the Einstein-Hilbert action, without reference to the conformal factor.
\end{abstract}

\maketitle

\section{Introduction}
\label{Sec:Intro}

In previous work I have shown how mechanical systems which have scaling symmetries can be reduced to descriptions which eliminate the scaling direction from the system. This reduction can be performed in both the (symplectic) Hamiltonian \cite{sloan2018dynamical} and Lagrangian \cite{sloan2023herglotz} settings, with the reduced descriptions taking the form of a contact Hamiltonian system and Herglotz Lagrangian system respectively. Such descriptions of cosmology eliminate the scale factor, but retain the Hubble parameter \cite{sloan2021new}. This has been of particular interest in investigating the dynamics of Einstein's equation near to cosmological singularities \cite{koslowski2018through,mercati2019through,sloan2019scalar,adamo2023gauge,hoffmann2024continuation} and in simple black hole models \cite{mercati2022traversing}. Here I show how the same results apply to field theories, and thus to general relativity (GR). In GR the symmetry is related to the choice of conformal factor, which can be eliminated from the system. Note that although the conformal factor is removed, the role of its derivatives remains. Thus the full dynamics of GR are encoded in action that depends only on a metric $\gamma$ of fixed determinant, at the cost of introducing action densities to the Lagrangian description. I will show in section \ref{Sec:GR} that the Herglotz Lagrangian for this system is
\be L^H = R(\gamma) - \frac{1}{6} \gamma_{ab} s^a s^b \ee
For now the key things to note are two: first, there is no dependence on any conformal factor, since the metric $\gamma$ is of fixed determinant (which could be chosen to be -1, for example) there is no factor of the square root of the determinant of the metric, in contrast to the Einstein-Hilbert action. Second, this is achieved through the introduction of a dependence of the Lagrangian on the action density $s^a$.

The paper is laid out to enable those familiar with the material covered in any section to jump directly to results of interest. In the following section, section \ref{Sec:Contact} I review the contact manifold description of classical mechanics in the Lagrangian and Hamiltonian settings. In section \ref{Sec:Overview} I provide an overview of previous results on dynamical similarity to bring the reader up to speed with the particle results, and in section \ref{Sec:Framework} I introduce the equivalent description of field theories, particularly the de Donder-Weyl form of Hamiltonian field theory. In sections \ref{Sec:Lag} and \ref{Sec:KHam} I present the main result of the reduction in the Lagrangian and Hamiltonian frameworks respectively. Section \ref{Sec:SF} provides a simple example of this in the context of two scalar fields in flat space, before moving on to section \ref{Sec:GR} which provides the main result in the context of general relativity, and examines the resulting dynamics of example systems. Finally in section \ref{Sec:Conclusions} I provide some discussion and further context for the work. 

A note on notation: I will be primarily using notation common to physics literature in which the usual Euler-Lagrange equations are:
\be \frac{\d}{\d t} \left(\frac{\partial L}{\partial \dot{q}} \right) - \frac{\partial L}{\partial q} = 0\ee
A more mathematically precise version of this is to consider flows on the tangent space $TC$, with the total derivative $\frac{\d}{\d t}$ being expressed in terms of Tulczyjew’s total derivative, $D_T$ which can be given in local coordinates on $T(TC)$ for a function $F:TC \rightarrow \mathbb{R}$:
\be D_T F(q,\dot{q}) = \frac{\partial F}{\partial q} \dot{q} + \frac{\partial F}{\partial \dot{q}} \ddot{q} \ee
In terms more familiar to physicists this would be called the total time derivative of $F$. 

With partial derivatives I will reserve the use of $\partial_a F$ for a function $F$ to indicate $\frac{\partial F}{\partial x^a}$ where $x^a$ are space-time coordinates. When partial derivatives are taken with respect to fields I will retain the more complete notation $\frac{\partial F}{\partial u}$. 

In a similar notational move, in order to reduce algebraic clutter which may prove confusing or distracting to the reader, I will use $F$ to express a function $F:M \rightarrow \mathbb{R}$ and its composition with a map between manifolds. So consider $\pi:N \rightarrow M $, then if $q$ is a point on $N$ mapped to a point $Q=\pi(q)$ under $\pi: N \rightarrow M$, I will denote $(F \circ \pi) (Q)$ as $F(q(Q))$ or simply $F(q)$.  

Following our notation conventions, the Euler-Lagrange equations for a field theory are
\be D_a \left(\frac{\partial L}{\partial (\partial_a q)} \right) - \frac{\partial L}{\partial q} = 0 \ee
where we follow the Einstein summation convention, and $D_a$ is the total derivative, which in common physics notation would be $D_a = \frac{\d }{\d x^a}$, so explicitly
\be D_a F(q, \partial_b q) = \frac{\partial F}{\partial q} \partial_a q + \frac{\partial F}{\partial(\partial_b q)} \partial_a (\partial_b q) \ee

To aid comprehension of a broader audience, unless otherwise stated I will work in Darboux coordinates. These are universally proven to exist in symplectic and contact systems, and thus always valid. The relationships between contact and symplectic systems are independent of this choice.

\section{Contact Lagrangian and Hamiltonian mechanics}
\label{Sec:Contact}

Here I will provide a brief overview of contact Hamiltonian and Lagrangian systems, adapted to the purposes of this paper. For a comprehensive review of the subject see \cite{deLeon:2020hnm}. Herglotz considered an extension of the usual Lagrangian description of mechanics to include the action itself as a dynamical quantity: 
\be L^H: TC \times \mathbb{R} \rightarrow \mathbb{R}  \text{ that is } L^H=L^H (q,\dot{q},s) \text{ where } \dot{s}=L^H \ee
Such descriptions are particularly successful in describing nonconservative systems, such as the damped harmonic oscillator. A Herglotz Lagrangian which depends on the action, $s$, matter variables $q$, and their first time derivatives $\dot{q}$ has equations of motion
\be \frac{\d}{\d t} \left(\frac{\partial L^H}{\partial \dot{q}} \right) - \frac{\partial L^H}{\partial \dot{q}} \frac{\partial L^H}{\partial s}  - \frac{\partial L^H}{\partial q} = 0 \quad \dot{s} = L^H \ee
where we see that the usual Euler-Lagrange equations are reproduced when the Lagrangian $L^H$ has no dependence on $s$. 

The nonconservative nature of this description is apparent if we consider the Lagrangian energy, $E_L = \dot{q}\frac{\partial L^H}{\partial \dot{q}}-L^H$ and the Lagrangian contact form, $\eta_L = -ds + \frac{\partial L^H}{\partial \dot{q}} dq$. In such case we find
\be \dot{E_L} = E_L \frac{\partial L^H}{\partial s} \quad \dot{\eta_L} = \frac{\partial L^H}{\partial s} \eta_L \ee
again recovering conservation in the case that the Lagrangian does not depend on the action. 

Following a Legendre transform, assuming that the Lagrangian is regular (i.e. $p = \frac{\partial L^H}{\partial \dot{q}}$ is invertible) we can form the contact Hamiltonian
\ba H^c &:& T^*C \times \mathbb{R} \rightarrow \mathbb{R} \text{ that is } H^c =H^c (s,p,q) \nonumber \\
    p &=& \frac{\partial L^H}{\partial \dot{q}} \quad H^c = p\dot{q} - L^H \ea
The contact Hamiltonian equations of motion, in Darboux coordinates where $\eta = -ds + p dq$ are \cite{bravetti2017contact,bravetti2017contact2}
\be \dot{q} = \frac{\partial H^c}{\partial p}  \quad \dot{p} = -p \frac{\partial H^c}{\partial s} - \frac{\partial H^c}{\partial q} \quad \dot{s} = p \frac{\partial H^c}{\partial \Pi} - H^c \ee
and again we see the nonconservative nature of our description in the evolution of the contact Hamiltonian and the contact form:
\be \dot{H^c} = - \frac{\partial H^c}{\partial s} H^c \quad \dot{\eta} = \frac{\partial H^c}{\partial s} \eta \ee 
Hence the Hamiltonian is not conserved, and independence of $H^c$ from a coordinate does not cause the imply the conservation of the corresponding momentum. However, in the case that $H^c$ is independent of a coordinate $q$ the corresponding momentum retains its proportionality to the contact Hamiltonian, i.e. 
\be\frac{\partial H^c}{\partial q}=0 \rightarrow \frac{\d}{\d t} \left(\frac{p}{H^c} \right) = 0 \ee
The non-conservation of the contact form is similarly indicative of the nonconservative nature of the framework: spaces of solutions as measured on contact phase space can focus (or disperse) along dynamical trajectories. Such behaviour has potential implications for statistical mechanics and the nature of the arrow of time \cite{bravetti2015liouville,gryb2024closed,gryb2021scale,gryb2021new}.

\section{Overview of Dynamical Similarity}
\label{Sec:Overview}

A thorough mathematical account of dynamical similarity is given in \cite{bravetti2023scaling}, see also \cite{azuaje2023scaling} for application to cosymplectic and cocontact systems. Here I will briefly recapitulate the program for unfamiliar readers, together with a worked example - the planar Kepler problem - to demonstrate the reduction of symplectic systems with scale symmetries to contact systems. Those already conversant with the concepts can safely skip ahead to the next section.

Let $\mathfrak{X}(M)$ denote the space of vector fields on the manifold $M$. A dynamical similarity, introduced in \cite{sloan2018dynamical}, of a vector field $X \in \mathfrak{X}(M)$ is a vector field $Y$ such that $[X,Y]=fX$ wherein $f: M\rightarrow \mathbb{R}$ is a function on $M$. In physical theories, we apply this to manifolds $M$ representing $TC$ or $T^* C$ - the tangent bundles or cotangent bundles over a configuration space $C$, i.e. the flows of Lagrangian or Hamiltonian dynamics. These similarities are of physical interest when the vector field $Y$ relates to a physically unobservable change, and the function $f$ is non-zero. In these cases we are mapping solutions to some equations of motion to other solutions related through this unobservable change - thus which are physically indistinguishable. The role of $f$ here is that is allows a different time parametrization of these solutions. 

A scaling symmetry is a particular type of dynamical similarity satisfying, for some $\Lambda \in \mathbb{R}$, 
\be \Lie_{Y_H} \omega = \omega \quad \Lie_{Y_H} H = \Lambda H \ee
Note that these are necessarily dynamical similarities as $[Y_H, X_H] = (\Lambda-1) X_H$. We call $\Lambda$ the degree of the scaling symmetry, also if $Y$ is a dynamical similarity, so trivially will be $\alpha Y$ for all $\alpha \in \mathbb{R}$, so the action of $Y$ on the symplectic structure is used to normalize $Y$. 

A familiar example of this is rescaling of the Kepler problem. The configuration space $C = \mathbb{R}^2 -\{0\}  \cong \mathbb{R}_+ \times S^2$ which we will parametrize by polar coordinate $(r,\theta)$. The equations of motion can be derived from the Lagrangian or Hamiltonian framework from 
\be L = \frac{\dot{r}^2}{2} +\frac{r^2 \dot{\theta}^2}{2} + \frac{1}{r} \quad \leftrightarrow  \quad H = \frac{P_r^2}{2}+\frac{P_\theta^2}{2r^2} - \frac{1}{r} \ee
In the Lagrangian case $TC = \mathbb{R}_+ \times S \times \mathbb{R}^2$ with coordinates $(r,\theta,\dot{r},\dot{\theta})$. The Euler-Lagrange equations give rise to the vector field $X_L \in \mathfrak{X} (M)$:
\be X_L = \dot{r} \partial_r + \dot{\theta} \partial_\theta + \left( r \dot{\theta}^2 - \frac{1}{r^2} \right) \partial_{\dot{r}} - \frac{2 \dot{r} \dot{\theta}}{r} \partial_{\dot{\theta}} \ee
From this the vector field 
\be Y_L = 2r \partial_r - \dot{r} \partial_{\dot{r}} - 3\dot{\theta} \partial_{\dot{\theta}} \ee
Satisfies $[Y_L,X_L]=-3X_L$. 
Physically it is informative to note that the vector field $Y_L$ is the projection onto $TC$ of the rescaling of time and space such that $r \rightarrow \lambda^2 r$ and $t \rightarrow \lambda^3 t$ - and the symmetry is analogous to Kepler's third law.

To find a Herglotz Lagrangian which reproduces our dynamics, we first note that by a change of coordinates on configuration space and time we can simplify $Y_L$. Letting 
\be \rho = 2 \log r \quad \d \tau= r^\frac{3}{2} \d t = e^{3\rho} \d t \ee
we can write $Y_L = \partial_\rho$, and the Lagrangian density becomes
\be L \d t = e^\rho \left(\frac{\rho'^2}{8} + \frac{\theta'^2}{2} + 1 \right) \d \tau \ee
and thus the dynamics are reproduced by the Herglotz Lagrangian
\be L^H = \frac{\theta'^2}{2} - \frac{s^2}{8} + 1 \quad s'=L^H \ee
From which we obtain the equation of motion for $\theta$:
\be \theta''+\frac{s \theta'}{4} = 0 \ee
Together with the fact that $L^H$ is the $\tau$ derivative of $s$, this reproduces the equations of motion in the original coordinates. 

In the Hamiltonian framework the flow on $T^*C$ is
\be X_H = P_r \partial_r + \frac{P_\theta}{r^2} \partial_\theta + \left(\frac{P_\theta^2}{r^3} - \frac{1}{r^2} \right) \partial_{P_r}  \ee
For which the corresponding dynamical similarity to $Y_L$ can be expressed:
\be Y_H = 2r \partial_r + P_\theta \partial_{P_\theta} -P_r \partial_{P_r} \ee
which again gives rise to $[Y_H, X_H]=-3X_H$. It's further a simple exercise to show the $Y_H$ is a scaling symmetry of degree -2.

Given a Hamiltonian scaling symmetry, $Y_H$ we can reduce the dynamical system to a contact Hamiltonian system. Consider Hamiltonian system consisting of the phase space $M_H =T^* C$, Hamiltonian function $H$ and symplectic form $\omega$. If there is a dynamical similarity $Y_H$, then we can choose a eigenfunction $\rho$ of $Y_H$ with eigenvalue 1, i.e. such that $Y_H (\rho) = \rho$. Then there exists a contact system with contact phase space $T^*C/Y_H$, contact Hamiltonian $H^c=H/\rho^N$ and contact form $\eta = \iota_{Y_H} \omega/\rho$ which reproduces the same dynamics as $T^*C, H, \omega$.

In the Kepler example we can use $\rho=\sqrt{r}$ as our scaling function, as $Y_H (\sqrt{r}) = \sqrt{r}$. We can then find the contact form in Darboux coordinates:
\be \eta = \frac{\iota_{Y_H} \omega}{\sqrt{r}} = -\d (2\sqrt{r} P_r) + \frac{P_\theta}{\sqrt{r}} \d \theta = -\d s + \Pi \d \theta \ee
and thus the contact Hamiltonian in these coordinates is
\be H = r H^c = \frac{r P_r^2}{2} + \frac{P_\theta^2}{2r} - 1 = \frac{s^2}{8} + \frac{\Pi^2}{2} - 1 \ee
Note that this contact Hamiltonian is the Legendre transform of the Herglotz Lagrangian found above. It gives rise to equations of motion
\be q' = \Pi \quad \Pi' = -\frac{\Pi s}{4} \quad s' = \frac{\Pi^2}{2}-\frac{s^2}{8} +1 \ee
in which a prime denotes a derivative with respect to $\tau$ which is related to $t$ by $\d t = r^\frac{3}{2} \d \tau$. It is a trivial exercise to show that in these are equivalent to the dynamics of the full Kepler system. 

\section{Mathematical Framework of Field Theories}
\label{Sec:Framework}

Our goal is to establish the nature of dynamical similarities in classical field theories. To do so, we must first show how classical field theories can be described in a Lagrangian and Hamiltonian framework that is suitable for expressing scaling symmetries directly. I will therefore provide a brief overview of ``frictional" field theories. These are the contact Lagrangian and k-symplectic Hamiltonian theories, which are the field theory counterparts to the Herglotz Lagrangian and contact Hamiltonian descriptions of mechanical systems. For an excellent mathematical introduction to k-symplectic theories and k-contact systems see \cite{gaset2020contact}, and discussions of \cite{de2015methods,Gnther1987ThePH,marrero2011kind,guerra2023more,de2023multicontact}. Here I will summarize the relevant results. In particular I will aim to describe these in a manner most accessible to those familiar with physics notations.

Let us consider a generalization of the action principle of the type proposed by Herglotz. In this we consider a Lagrangian $L^H$ which depends not only on values of fields which exist on our spacetime manifold, but also upon an action density $s^a$, which is related to the Lagrangian $L^H$ by $\partial_a s^a = L^H$. Note that this is a generalization to field theory of the usual particle case in which $L=\frac{\d S}{\d t}$. It is common to express this relationship in integral form:
\be S = \int_M L \, \mathrm{dVol} = \int_{\partial M} n_a s^a \d \sigma \ee
where $\partial M$ is the boundary of $M$, with volume element $\d \sigma$ and $n_a$ the unit normal. In many cases we will consider manifolds of the form $\Sigma \times I$ in which $I \subset \mathbb{R}$ is a time interval, and $\Sigma$ is a spatial slice. In such cases, in coordinates adapted to this splitting
\be S = \int_{\Sigma_f} s^t \d \sigma - \int_{\Sigma_i} s^t \mathrm{dVol}_\Sigma \ee
Per convention, we will fix our fields on the boundary, and extremize the action within. The critical points are solution to the Herglotz equations:
\be D_a \left(\frac{\partial L}{\partial (\partial_a q)} \right) - \frac{\partial L}{\partial (\partial_a q)} \frac{\partial L}{\partial s^a} - \frac{\partial L}{\partial q} = 0 \ee

For the Hamiltonian description of field theories, our starting point will be the de Donder-Weyl description of an autonomous field theory in terms of k-symplectic structures. The goal of this is to create a covariant formulation of Hamiltonian field theory. As such as well as replacing time derivatives of fields with momenta, we also replace spatial derivatives.  Therefore for each field, at a given point in a $k$-dimensional space-time we will have $k+1$ values consisting of the value of the field itself together with its momenta. 

Our basic elements will consist of a $n$ fields, $\psi^a$ on a $k$-dimensional space-time manifold $M_s$, with $k$ copies of the cotangent bundle $T^* M$, $k$ closed 2-forms $\omega^a =\d \theta^a$, and an $nk$ dimensional tangent distribution, $V$ (this defines the space of momenta for our fields). Our k-symplectic system consists then of $M$, the $k(n+1)$ dimensional manifold of the $n$ fields and each of their $k$-momenta. Fortunately we are guaranteed the existence of Darboux coordinates $q_i, p_i^a$, and in these we can express our structures
\be \omega^a = \d p^a \wedge \d q \quad V = \mathrm{Span} \left(\frac{\partial}{\partial p_i^a} \right) \ee
Our Hamiltonian will be a function of the $q$ and $p^a$. Note that this differs from the usual Hamiltonian formulation of field theory in which only the `time' momenta are used, and we retain spatial derivatives of the $q$s. The equations of motion for our system are then obtained from
\be \iota_X^a \omega^a = -\d H \ee
which determines the Hamiltonian vector fields $X^a$. In Darboux coordinates these become
\be \partial_a q = \frac{\partial H}{\partial p^a} \quad \partial_a p^a = -\frac{\partial H}{\partial q} \ee
As is apparent, these closely resemble the usual Hamilton's equations in Darboux coordinates, and would reduce to them in the case of a 1-dimensional space-time. We note here that there is a subtle issue regarding the integrablity of such systems, in particular the existence of of a field $q$ whose derivatives are of this form. For a discussion of such issues see \cite{Gnther1987ThePH,de2015methods}. Henceforth we will work with systems for which this condition holds, both in the Lagrangian and Hamiltonian settings.

A k-contact system extends the idea of a k-symplectic system by introducing a further $k$ dimensions to the manifold. For our purposes it will suffice to focus on a simple case. Consider $M=M_s \times \mathbb{R}^k$, where $M_s$ is a k-symplectic manifold, with symplectic potentials $\theta^a = p^a dq$. We can extend these to contact forms on $M$ by taking $\eta^a = -ds^a + \theta^a$, where $s^a$ are Darboux coordinates on $\mathbb{R}^k$. A k-contact Hamiltonian is a function $H^c$, with equations of motion for fields on $M$ being
\be \partial_a q = \frac{\partial H^c}{\partial p^a} \quad \partial_a p^a = -\frac{\partial H^c}{\partial s^a} p^a - \frac{\partial H^c}{\partial q} \quad \partial_a s^a = p^a \frac{\partial H^c}{\partial p^a} - H^c \ee

As would be expected, subject to conditions of regularity, a k-contact Hamiltonian can be obtained following a Legendre transform of a Herglotz Lagrangian, where $p^a = \frac{\partial L^H}{\partial (\partial_a q)}$ and $H^c$ = $p^a \partial_a q - L^H$.

\section{Scaling Symmetries: Lagrangian Setup}
\label{Sec:Lag}
Now that I have introduced the necessary frictional field theories, I will show how the existence of a scaling symmetry can be used to reduce non-frictional theories to their frictional counterparts. First I will do this in the Lagrangian framework.

Consider a Lagrangian defined on the first jet over a configuration space $C$ of fields on a space-time manifold $M$:  $L:J^1 C \rightarrow \mathbb{R}$ on which there exists a global scaling symmetry, $\D: \in T(J^1 C): \Lie_\D L=L$. Then we can pick coordinates such that $\D=\partial_x$, and foliate $J^1 C$ by $x=\mathrm{Constant}$ slices: $J^1 C=\mathbb{R} \times K$. In such a decomposition, 
\be L(x,u_a,q,v_a) = e^x F(u_a,q,v_a) \quad \mathrm{where} \quad u_a = \partial_a x \quad v_a = \partial_a q \ee
Our goal at this point is to see that there is a description of the same system which can be expressed without reference to $x$, and thus is a more compact description. This will turn out to be a Herglotz Lagrangian which we will explicitly construct below. 

The Euler-Lagrange equation for $q$ is then
\be D_a \left(\frac{\partial F}{\partial v_a}\right) + u_a \frac{\partial F}{\partial v_a} - \frac{\partial F}{\partial q} = 0 \ee
and the equation for $x$ is 
\be D_a \left(\frac{\partial F}{\partial u_a} \right) + u_a \frac{\partial F}{\partial u_a} - F =0 \label{MotivHerg} \ee
The second of these, equation \ref{MotivHerg}, provides motivation for considering $\frac{\partial F}{\partial u_a}$ to be an action density for a Herglotz Lagrangian consisting of the remaining terms. Let us define $s^a(u_b, q,v_b) = \frac{\partial F}{\partial u_a}$. Let us further posit that this relationship can be inverted to determine $u_a (s^b, v_b, q)$. Note that this is not a particularly big demand. E.g. In the case of Lagrangians quadratic in velocities 
\be F = g^{ab} (q) u_a u_b + G(q,v_a) \rightarrow u_a = g_{ab} (q) s^b \ee
We can rearrange the Euler-Lagrange equation for $x$ to show
\be \partial_a s^a = F - u_a s^a \ee
which shows that we can indeed treat $s^a$ as an action density with Herglotz Lagrangian density
\be L^H (s^a, q, v_a) = F(u_a (s^b,q,v_b), q, v_a) - u_a (s^b, q, v_b) s^a \ee
where I have kept the dependences explicit since we're going to have to be careful with partial derivatives. Let us denote a partial derivative on the space of $s^a, q, v_a$ by $\tilde{\partial}$. Having used the above definitions to motivate considering $L^H$ to be a Herglotz Lagrangian, let us now show that the equations of motion for the other fields are indeed reproduced by the Herglotz equations for $L^H$. 

Then the Herglotz-Lagrange equation for $q$ is then 
\be D_a \left(\frac{\tilde{\partial} L^H}{\tilde{\partial} v_a} \right) - \frac{\tilde{\partial} L^H}{\tilde{\partial} v_a} \frac{\tilde{\partial} L^H}{\tilde{\partial} s^a} - \frac{\tilde{\partial} L^H}{\tilde{\partial} q} = 0 \ee
we can evaluate each term here in turn:
\be \frac{\tilde{\partial} L^H}{\tilde{\partial} v_a} = \frac{\partial F}{\partial u_b} \frac{\partial u_b}{\partial v_a} + \frac{\partial F}{\partial v_a}  - \frac{\partial u_b}{\partial v_a} s^b = \frac{\partial F}{\partial v_a} \ee
since the first and third terms cancel, as $s^b = \frac{\partial F}{\partial u_b}$. 
\be \frac{\tilde{\partial} L^H}{\tilde{\partial} s^a} = \frac{\partial F}{\partial u_b} \frac{\partial u_b}{\partial s_a} - u_a = \frac{\partial u_b}{\partial s^a} s^b = -u_a \ee
and
\be \frac{\tilde{\partial} L^H}{\tilde{\partial} q} = \frac{\partial F}{\partial u_b} \frac{\partial u_b}{\partial q} + \frac{\partial F}{\partial q} - \frac{\partial u_b}{\partial q} s^b = \frac{\partial F}{\partial q} \ee
Having now amassed all the pieces we need, we see that the Herglotz-Lagrange equation for $q$ is thus:
\be D_a \left(\frac{\partial F}{\partial v_a} \right) + u_a \frac{\partial F}{\partial v_a} - \frac{\partial F}{\partial q} = 0 \ee
which is exactly the Euler-Lagrange equation for $q$ that we had derived earlier. Thus the Herglotz Lagrangian density $L^H$ reproduces the same dynamics as derived from $L$.

Let us summarize this result: Given a Lagrangian with a scaling symmetry, we can exploit the scaling symmetry to find a Herglotz Lagrangian which reproduces exactly the dynamics of the fields yet is more parsimonious in its setting, depending only on the fields unaffected by the scaling symmetry. The role of the derivatives of the field $\rho$ has been taken on by the action density $s^a$.

It is clear that this process can be inverted, so beginning with a Herglotz Lagrangian $L^H(s^a, q, v_a)$ we can define $L(x, u_a, q, v_a)=e^x(L^H + s^a u_a)$ where $u_a =-\frac{\partial L^H}{\partial s^a}$. A direct calculation of the Euler-Lagrange equations for $x$ is outlined:
\ba \frac{\partial L}{\partial x} &=& e^x \left( L^H + u_a s^a + \frac{\partial L^H}{\partial s^a} \frac{\partial s^a}{\partial \rho} +u_a \frac{\partial s^a}{\partial \rho} \right) = e^x (L^H + s^a u_a)  \nonumber \\ 
\frac{\partial L^H}{\partial u_a} &=& e^x \left( \frac{\partial L^H}{\partial s} \frac{\partial s}{\partial u_a} + s + u_a \frac{\partial s}{\partial u_a} \right) = e^x s^a \nonumber \\ 
&\rightarrow& \partial_a s^a = L^H \ea
and the Euler-Lagrange equation for q is
\ba \frac{\partial L}{\partial q} &=& e^x \left( \frac{\partial L^H}{\partial q} + \frac{\partial L^H}{\partial s^a} \frac{\partial s^a}{\partial q} + u_a \frac{\partial s^a}{\partial q} \right) = e^x \frac{\partial L^H}{\partial q} \nonumber \\
\frac{\partial L}{\partial v_a} &=& e^x \left(\frac{\partial L^H}{\partial v_a} + \frac{\partial L^H}{\partial s^b}{\partial s^b}{\partial v_a} + u_b \frac{\partial s^b}{\partial v_a} \right) = e^x \frac{\partial L^H}{\partial u_a} \nonumber \\
&\rightarrow& D_a\left(\frac{\partial L^H}{\partial v_a}\right) - \frac{\partial L^H}{\partial s^a} \frac{\partial L^H}{\partial v_a} - \frac{\partial L^H}{\partial q} = 0
\ea

From the construction of our Lagrangian it is obvious that there exists a scaling symmetry, $\D=\frac{\partial}{\partial x}$ on this system. Thus we see that we can exchange a Lagrangian with a scaling symmetry for a Herglotz Lagrangian on the reduced space, and conversely extend a Herglotz Lagrangian through embedding into a larger space and the introduction of a new scaling symmetry. 

A note on boundary terms: Consider the case where a Herglotz Lagrangian found in this was has a total derivative term: $L^H_2 = L^H_1 + \partial_a X$ for some general $X$. In this case we can scale-extend the Herglotz Lagrangians, so $L_2 = e^A F_2 = e^A (F_1 + \partial_a X)$. Then at the cost of a boundary term, we can eliminate $\partial_a X$ to find 
\be L_2 = e^A (F_1 + X \partial_a A) \ee
and finding the equivalent Herglotz-Lagrange scale reduction we see that $s_2^a =s_1^a + X$, i.e.
\be L^H_2 = F_1 + X \partial_a A - s_2^a \partial_a A = L^H_1 \ee
and thus we can eliminate the total derivative term from our Herglotz Lagrangian through replacing $s^a$ by $s^a + X$, as we would expect from $s^a$ being an action density. This will be of particular importance when dealing with general relativity where the action is technically second order, but we will eliminate second derivatives from our description by replacing them with first derivatives and a boundary term. 

\section{Scaling Symmetries: k-symplectic Hamiltonian Setup}
\label{Sec:KHam}

I will now show how the same scaling symmetries can be used directly in the Hamiltonian framework to reduce a de Donder-Weyl Hamiltonian system to a k-contact system.

Let us consider a k-symplectic Hamiltonian system $(M_H,\omega^a, H)$. Our goal now is to examine the case in which there is a dynamical similarity present in these fields. The strategy will run parallel to that employed in the particle case. On identifying a dynamical similarity, we will choose coordinates such that on the configuration space translation along this coordinate will correspond to a rescaling. We will then use the dynamical similarity to eliminate this coordinate. On space of momenta this will leave $k$ momenta which are no longer related to a field, and upon rescaling these will become our Reeb fields. Thus we will move our setting from $M_H$ of dimension $D=n(k+1)$ to $M_c$ of dimension $(n-1)(k+1)+k=D-1$. 

Let us begin with our configuration space of fields $Q$ and let $M_H = \oplus^k T^* Q$. Recall that the equations of motion for fields are determined by elements $X^a \in TM_H$ satisfying
\be \label{KsympEoM} \iota_{X_a} \omega^a = -\d H \ee
Let us suppose that there exists a scaling symmetry $\D$ such that 
\be \Lie_\D H = H \quad \Lie_\D \omega^a = \omega^a \ee
In parallel with the particle case we will choose coordinates on $Q$ such that the projection of $\D$ onto $Q$ is $\partial_x$, and work in Darboux coordinates. To satisfy the definition of a dynamical similarity, requires $\D  = \partial_x + P^a_i \partial_{P^a_i}$. Hence we can decompose $Q$ such that
\be Q = \mathbb{R} \times Q_c \rightarrow M_H = \oplus^k T^*\mathbb{R} \times \oplus^k T^* Q_c =\mathbb{R}\times M_c \ee 
Here we see how our manifold will break into a contact configuration space $Q_c$, the Whitney sum of tangent spaces over $Q_c$ (corresponding to contact momenta) and $k$ copies of $\mathbb{R}$, which will become the Reeb fields. 

In our decomposition, $H=e^x H^c$ and $\omega^a = \d (e^x \eta^a)$ where $\Lie_D H^c = 0 = \Lie_\D \eta$, hence both $H^c$ and $\eta$ are independent of $x$. Thus 
\be \d H = e^x (\d H^c + H^c \d x) \quad \omega^a = e^x (\d x \wedge \eta^a + \d \eta^a) \ee
From the equation \ref{KsympEoM}, we see
\be \iota_{X_a} \eta^a = \iota_{X_a} e^{-x} \iota_\D \omega^a = -e^{-x} \iota_\D \iota_{X_a} \omega^a = e^{-x} \iota_\D dH = e^{-x} \Lie_\D H = H^c \ee
and hence $(M_c ,H^c, \eta^a)$ is a k-contact Hamiltonian system, with the restriction of $X_a$ to $M_c$ being the Hamiltonian vector fields of $M_c$.  

A perhaps more illuminating understanding can be obtained by considering Lagrangians of the type examined in the previous section. The k-momenta are then
\be \frac{\partial L}{\partial u_a} = e^x \frac{\partial F}{\partial u_a} = e^x s^a \quad \frac{\partial L}{\partial v_a} = e^x \frac{\partial F}{\partial v_a} = e^x B^a \ee
In terms of these variables, the symplectic structures are then
\be \omega^a = e^x (\d s^a \wedge \d x + B^a \d x \wedge \d q + \d B^a \wedge \d q) \label{HamSym} \ee	
and the Hamiltonian is 
\be H = e^x (s^a u_a + B^a v_a - L) =e^x H^c \ee
At this point let us note that we did not require the existence of a Lagrangian $L$ to form $H$, all we have required is the (guaranteed) existence of Darboux coordinates, wherein 
\be \omega^a = \d P^a \wedge \d q \ee
and thus letting $P_x^a=e^x s^a$ and $P_q^a = e^x B^a$ reproduces the necessary form of the symplectic structures in equation \ref{HamSym}. Then the requirement that there exists a dynamical similarity $\D$ is sufficient to ensure that we can write $H=e^x H^c (s^a, q, B^a)$. 

The equations of motion expressed in these variables, obtained from $\iota_{X_a} \omega^a = -\d H$ are:
\be \partial_a s^a = B^a \partial_a q - H^c \quad \partial_a x = \frac{\partial H^c}{\partial s^a} \quad \partial_a q = \frac{\partial H^c}{\partial B^a} \quad \partial_a B^a = -\frac{\partial H^c}{\partial q} - B^a \frac{\partial H^c}{\partial s^a} \ee
There are the k-contact Hamiltonian equations for the contact Hamiltonian $H^c=e^{-x} H$, with contact forms 
\be \eta^a = e^{-x} \iota_\D \omega = -\d s^a + B^a \d q \ee
This is confirmed by considering the Legendre transform of the Herglotz Lagrangian:
\be L = F-u_a s^a \rightarrow H^c = v_a \frac{\partial F}{\partial v_a} - L = e^{-x} H \ee
Hence we see that the process of contact reduction can be performed either on the Lagrangian or Hamiltonian description of a system, and we recover the same contact Hamiltonian or Herglotz Lagrangian description through a Legendre transform. This is illustrated in figure \ref{Commutative}. 
\begin{center}
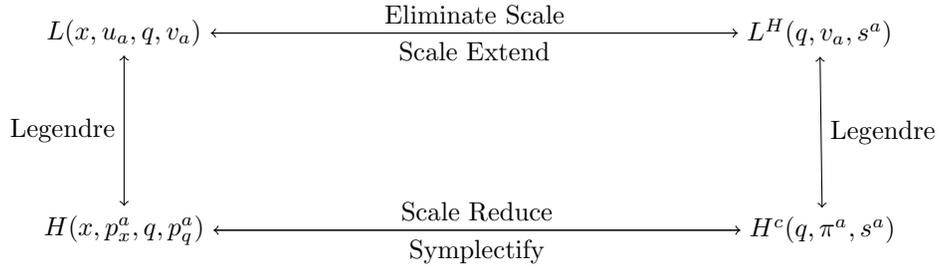
\begin{figure}[ht]
\begin{tikzpicture}[
  node distance=2cm and 7cm,
	auto]
  \node (L) {$L(x,u_a,q,v_a)$};
  \node (LH)[right= of L]{$L^H(q,v_a,s^a)$ };
  \node (H)[below= of L]{$H(x,p_x^a,q,p_q^a)$};
  \node (HC)[right= of H]{$H^c(q,\pi^a,s^a)$};
  \draw[<->] (L) to node [swap] {Legendre} (H);
  \draw[<->] (LH) to node {Legendre} (HC);
  \draw[->] (L) to node {Eliminate Scale} (LH);
  \draw[->] (LH) to node {Scale Extend} (L);
  \draw[->][yshift=1cm] (H) to node {Scale Reduce} (HC);
  \draw[->] (HC) to node {Symplectify} (H);
\end{tikzpicture}
\caption{Commutative diagram, showing the relationship of the Lagrangian, Herglotz Lagrangian, symplectic Hamiltonian and contact Hamiltonian frameworks.}
\label{Commutative}
\end{figure}
\end{center}
The process of symplectifying a contact Hamiltonian is relatively simple in this context and gives rise to a Hamiltonian system with a dynamical similarity. Given a contact Hamiltonian system $(M_c, H^c, \eta^a)$, let us work in Darboux coordinates such that $\eta^a = -\d s^a + B^a \d q$, and extend our phase space by letting $M=\mathbb{R}\times M_c$, with $x$ the coordinate along $\mathbb{R}$. Then if we let $\omega^a = \d (e^x \eta^a)$ we recover the symplectic structure of equation \ref{HamSym}, from which we know the equations of motion given by $H=e^x H^c$ match those of the k-contact system. Thus we can always symplectify a k-contact Hamiltonian system to produce an equivalent k-symplectic Hamiltonian system with dynamical similarity.

It is informative to note that our approach has necessitated the use of a k-symplectic Hamiltonian system for reduction, rather than a regular Hamiltonian system. This is well illustrated by the fact that our k-contact Hamiltonian is a function of the k-momenta and the fields, but does not contain any explicit derivative terms. Had we instead followed the more familiar (symplectic) formalism in which only momenta corresponding to the time derivatives of the fields were involved, then our Hamiltonian would have been the integral over a spatial slice of a function of these momenta, fields, and the spatial derivatives of the fields. If the Hamiltonian contained spatial derivatives of the field $x$, the usual process would have been to vary the Hamiltonian, which would have given rise to terms for the form $\delta \partial_a x$. In the symplectic system we would here integrate by parts to recover a term proportional to $\delta x$. In the contact case this would mean that we had failed in our  goal of eliminating $x$ from the set of dynamical variables.

\section{Example: Two scalar fields in flat space}
\label{Sec:SF}

In order provide a clear application each of the constructions that I have introduced in the previous sections, I will demonstrate the use of each in the case of a simple toy model, a two-dimensional real scalar field on a Minkowski background. This model is chosen due to its relative simplicity being a close parallel to a two dimensional harmonic oscillator. 

Consider a two-dimensional scalar field, $C=\mathbb{R}^2$, with Lagrangian $L$. The action is
\be S = \frac{1}{2}  \int \eta^{ab} (\partial_a \phi \partial_b \phi +\partial_a \psi \partial_b \psi) - m^2 (\phi^2 + \psi^2) d^4x \ee
wherein $\eta^{ab}$ is the metric on Minkowski space, i.e. $\eta=\rm{Diag}(-1,1,1,1)$. The first thing to establish is that there exists a scaling symmetry in this system. To see this we perform a change of coordinates for the field to mimic exponential polar coordinates. Doing so allows us to express the Lagrangian density in a manner compatible with our analysis. Let
\be \left(\begin{array}{c}\phi\\ \psi \end{array}\right) = e^\frac{\rho}{2} \left( \begin{array}{c}
\cos\theta \\ \sin \theta
\end{array}\right) \ee 
Then we see 
\be S =\int e^\rho \left(\frac{1}{8} \eta^{ab} \partial_a \rho \partial_b \rho +  \frac{1}{2} \eta^{ab} \partial_a \theta \partial_b \theta  - \frac{m^2}{2} \right) \d^4x \label{2SFLag} \ee
It is clear from this that $\D = \frac{\partial}{\partial \rho}$ is a scaling symmetry in this coordinate description. 

The Euler-Lagrange equations for our two scalar field system give
\ba \frac{\eta^{ab}\partial_a \partial_b \rho}{2} + \frac{\eta^{ab} \partial_a \rho \partial_b \rho}{4} - \eta^{ab} \partial_a \theta \partial_b \theta +m^2 = 0 \nonumber \\
\eta^{ab} \partial_a \partial_b \theta + \eta^{ab} \partial_a \theta \partial_b \rho = 0 \ea 
The Euler-Lagrange equations are thus the Klein-Gordon equation for the fields, $(\Box+m^2)\psi=0=(\Box+m^2) \phi$. 

The next task is to exploit the scaling symmetry under translation of $\rho$. It is informative to note here that I have chosen a field parametrization such that a translation of $\rho$ does not affect anything other than the pre-factor of the Lagrangian - other choices of fields (such as $R \propto \log \rho$) would require us to also rescale the kinetic terms within the Lagrangian itself. This is certainly possible, and may in some circumstances be enlightening as to the physical nature of the symmetry itself. In the analogous particle case - the harmonic oscillator - this corresponds to the distance of the particle from the central point of the potential. However, my goal here is to elucidate the mathematical construction in the simplest manner possible, hence the use of $\rho$ which keeps the algebra relatively simple and highlights the process of constructing the reduced systems.

To arrive at the Herglotz description, we note here that per our procedure $s^a = \frac{\eta^{ab} \partial_b \rho}{4}$ and hence the Herglotz Lagrangian is
\be L^H = \frac{1}{2}\eta^{ab} \partial_a \theta \partial_b \theta - 2 \eta_{ab} s^a s^b - \frac{m^2}{2} \label{2SFHer} \ee
It is a simple calculation so show that the  Euler-Lagrange equation for $\rho$ is reproduced by the fact that $s^a$ is the action density for this Lagrangian (i.e. $\partial_a s^a = L^H$). This exemplifies one of reasons that such a simplification can exist: the action for our system plays a dynamical role, rather than being simply a quantity that is extremized. The fact that $s^a$ is an action density is the equivalent to a dynamical equation in the Lagrangian representation in its own right. Further, the Herglotz-Lagrange equation for $\theta$ is
\ba D_a \left(\frac{\partial L^H}{\partial (\partial_a \theta)} \right) - \frac{\partial L^H}{\partial (\partial_a \theta)} \frac{\partial L^H}{\partial s^a} - \frac{\partial L^H}{\partial \theta}= 0 \nonumber \\
\eta^{ab} \partial_a \partial_b \theta + 4 s^a \partial_a \theta = 0 \ea
Thus we have arrived at the same equations of motion from our Herglotz system as we did from the original Lagrangian. However, while we have retained the action density, $s^a$, we make no reference to the original variable $\rho$ anywhere in this system. Nonetheless, the dynamics of the fields are reproduced exactly. 

To illustrate the reverse process, let us take the Herglotz-Lagrange equation \ref{2SFHer} as our starting point. Following the process given in section \ref{Sec:Lag} we introduce the field $\rho$ where 
\be \partial_a \rho =  -\frac{\partial L^H}{\partial s^a} = 4 \eta_{ab} s^b \ee
and thus our Lagrangian is 
\be L = e^\rho (L^H + s^a \partial_a \rho) = e^\rho \left(\frac{1}{2} \eta^{ab}\partial_a \theta \partial_b \theta +\frac{1}{8} \eta^{ab} \partial_a \rho \partial_b \rho -\frac{m^2}{2} \right) \ee
which exactly reproduces equation \ref{2SFLag} as expected.

To obtain the the k-symplectic Hamiltonian, we perform a Legendre transform of equation \ref{2SFLag}. Our momenta are then
\be P_\rho^a = \frac{\partial L}{\partial (\partial_a \rho)} = \frac{1}{4} e^\rho \eta^{ab} \partial_b \rho \quad P_\theta^a = \frac{\partial L}{\partial (\partial_a \theta)} = e^\rho \eta^{ab} \partial_b \theta \label{2SFLeg} \ee
and hence we arrive at the k-symplectic Hamiltonian system
\be H = 2 e^{-\rho} \eta_{ab} P_\rho^a P_\rho^b + \frac{1}{2} e^{-\rho} \eta_{ab} P_\theta^a P_\theta^b + e^\rho \frac{m^2}{2} \quad \omega^a = \d P_\rho^a \wedge \d \rho + \d P_\theta^a \wedge \d \theta  \ee
The equations of motion for this system are unsurprisingly equivalent to those of the original Lagrangian system. The two equations for the evolution of the configuration space variables $\rho$ and $\theta$ are equivalent to the definitions of the momenta given by the Legendre transform of equation \ref{2SFLeg}.
\be \partial_a \theta = e^{-\rho} \eta_{ab} P_\theta^b \quad \partial_a \rho = 4 e^{-\rho} \eta_{ab} P_\rho^b \ee
and the equations of motion for the momenta reproduce the Euler-Lagrange equations when given the above:
\ba \partial_a P_\theta^a &=& 0 \rightarrow \eta^{ab} \partial_a \theta \partial_b \theta + \eta^{ab} \partial_a \theta \partial_b \rho = 0 \nonumber \\
\partial_a P_\rho^a &=& 2 e^{-\rho} \eta_{ab}P_\rho^a P_\rho^b + \frac{1}{2} e^{-\rho} \eta_{ab} P_\theta^a P_\theta^b - e^\rho \frac{m^2}{2} \nonumber \\ 
					&\rightarrow & 
\eta^{ab}\partial_a \partial_b \rho + \frac{1}{2} \eta^{ab} \partial_a \rho \partial_b \rho - 2\eta^{ab} \partial_a \theta \partial_b \theta + m^2 = 0 \ea
Our k-symplectic Hamitlonian system has a dynamical similarity resulting from the scaling symmetry:
\be \D = \frac{\partial}{\partial \rho} + P_\rho^a \frac{\partial}{\partial P_\rho^a} + P_\theta^a \frac{\partial}{\partial P_\theta^a} \ee
Let us now follow the previous section to reveal the k-contact Hamiltonian system. Since we began with a description of the system in Darboux coordinates for the k-symplectic manifold, it is unsurprising that we recover Darboux coordinates for the k-contact manifold. The dynamical similarity $\D$ is quite simple in this form, being a combination of rescaling the momenta and translating the scaling field $\rho$:
\be \eta^a = -\d s^a + \pi^a \d \theta \quad H^c = 2 \eta_{ab} s^a s^b +\frac{1}{2} \eta_{ab} \pi^a \pi^b +\frac{m^2}{2} \ee

As prescribed above we note that the k-contact structures are $\eta^a = e^{-\rho} \iota_\D \omega^a$. 
The contact Hamiltonian equations for our new system are then:
\ba \partial_a \theta &=& \frac{\partial H^c}{\partial \pi^a} \rightarrow \pi^a = \eta^{ab} \partial_b \theta \nonumber \\
    \partial_a \pi^a &=& -\frac{\partial H^c}{\partial \theta} - \pi^a \frac{\partial H^c}{\partial s^a} \rightarrow \eta^{ab}\partial_a \partial_b \theta + 4 s^a \partial_a \theta \nonumber \\ 
    \partial_a s^a &=& \pi^a \partial_a \theta - H^c = \frac{1}{2} \eta^{ab} \partial_a \theta \partial_b \theta - 2\eta_{ab} s^a s^b -\frac{m^2}{2} =L^H
    \ea
    
Let us here take stock of the constructions that this example highlights. First we have observed the system of two scalar fields on Minkowski space in both the Lagrangian and k-symplectic description. Through a choice of parametrization of the fields, much akin to polar coordinates with exponentiated radial direction, we have seen that there is a simple scaling symmetry. This scaling symmetry was then used in both cases to reveal the Herglotz-Lagrangian and k-contact field descriptions respectively, in which the radial direction was eliminated. I have then shown that the two cases the dynamics reproduced those of the full system, without reference to the radial direction, thus reducing the number of degrees of freedom necessary to describe the complete system. Although this system was relatively simple, it sufficed in demonstrating the reduction. In the next section, I will tackle a more physically interesting system - that of general relativity. 

\section{Application to General Relativity}
\label{Sec:GR}

The general theory of relativity describes the dynamics of space-time as a physical entity in its own right. The dynamical content - the Einstein Field Equations (EFE) can be derived from an action principle due to Einstein and Hilbert. This action depends on the metric $g$ and its derivatives. 

The Einstein-Hilbert action is 
\be S = \int \sqrt{-g} R \, \d^4x \ee
Let $g_{ab} = e^\phi \gamma_{ab}$ where $\phi(x)$ is a conformal factor, and $\gamma_{ab}$ has determinant -1, hence $\sqrt{-g}=e^{2\phi}$. A standard result is that the Ricci scalar can be decomposed:
\be R(g) = e^{-\phi} \left( R(\gamma) - 3 \gamma^{ab} \nabla_a \nabla_b \phi - \frac{3}{2} \gamma^{ab} \partial_a \phi \partial_b \phi \right) \ee
and thus the Einstein-Hilbert action becomes, up to a boundary term,
\be \label{EHPhi} S = \int e^\phi \left(R(\gamma) +\frac{3}{2} \gamma^{ab} \partial_a \phi \partial_b \phi \right) \d^4x \ee
and thus is of the form described in the section \ref{Sec:Lag}.

This observation leads us to identify the action density:
\be s^a = 3 \gamma^{ab} \partial_b \phi \ee
with Herglotz Lagrangian density
\be \label{LHGR} L^H = R(\gamma) - \frac{1}{6} \gamma_{ab} s^a s^b \ee
From our previous results, we know that will reproduce that dynamics of GR without ever including the conformal factor. The role of the derivatives of the conformal factor in the Einstein-Hilbert action is reproduced by the action densities. For completeness, a direct demonstration of the equivalence is also given in appendix \ref{Sec:Appendix}. 

Restricting ourselves to metrics which are homogeneous we can find the cosmological sector of GR, in particular the class A Bianchi models. Let us suppose that our space-time manifold is decomposed as $I \times \Sigma$ where $I \in \mathbb{R}$ is a time interval and $\Sigma$ is a maximally symmetric three-dimensional manifold with one-forms $\sigma_1, \sigma_2, \sigma_3$, corresponding to the three Killing vectors. We can thus express the metric as
\be \d s^2 = -\d t^2 + a_1^2 \sigma_1^2 + a_2^2 \sigma_2^2 +a_3^2 \sigma_3^2  \ee
Through a choice of variables $\phi = \frac{2}{3} \log( a_1 a_2 a_3), \d t = e^\frac{\phi}{2} \d \tau$ we can bring this metric into the form in which the conformal factor becomes apparent:
\be \d s^2 = e^\phi (-\d \tau^2 + e^{2\Gamma_i} \sigma_i^2) \ee
where $\Gamma_i =  \log(a_i)-log(\phi)$. There is a redundancy here as the sum of the $\gamma_i$ can be chosen to be 1. We can thus write the metric in Misner coordinates where $\Gamma_i$ is expressed in terms of two coordinates $x$ and $y$:
\be \d s^2 = e^\phi \left(-\d \tau + e^{y-\frac{x}{\sqrt{3}}} \sigma_1^2 + e^{-y-\frac{x}{\sqrt{3}}} \sigma_2^2 + e^{\frac{2x}{\sqrt{3}}} \sigma_3^2 \right) \ee
The unit determinant metric consists of the above without the conformal factor (i.e. setting $\phi$ to zero). Applying this we can find the Lagrangian of equation \ref{LHGR} for such systems and thus find the action, after integrating over a suitable fiducial cell or spatial slice, depending on compactness (the volume of which we set to 1 for simplicity):
\be S_{\mathrm{Bianchi}} = \int \left( \frac{\dot{x}^2}{2} + \frac{\dot{y}^2}{2} - {}^3R(x,y) + \frac{s_\tau^2}{6} \right) \d \tau \ee
which precisely reproduces the action presented in \cite{sloan2023herglotz}. In this note that we have set the spatial action densities to zero to preserve homogeneity, and ${}^3R$ is the Ricci scalar of the spatial three metric \emph{with unit determinant}, often referred to as the `shape potential'. 

Let us now move beyond the homogeneous sector. A full examination of the contact Hamiltonian formulation of general relativity is beyond the scope of this paper. However the general program can be demonstrated by application to the Lem\^aitre-Tolmann-Bondi system \cite{enqvist2008lemaitre}. Here we have an inhomogeneous, locally isotropic space-time with metric
\be \d s^2 = -\d \tau^2 + \alpha^2(\tau, r) \d r^2 + \beta^2(\tau,r) (\d \theta^2 + \sin^2 \theta \d \chi ) \ee
Let us form two new functions $A$ and $B$ where $A= \frac{2}{3} \log(\alpha) +\frac{4}{3} \log(\beta)$ and $B = \frac{2}{3} \log(\beta) - \frac{2}{3} \log(\alpha)$. Choosing our lapse such that $\d \tau = e^{A} \d t$ we can rewrite this metric as 
\be \d s^2 = -e^A \d t^2 + e^{A-2B} \d r^2 + e^{A+B} (\d \theta^2 + \sin^2 \theta \d \chi^2) \ee
where we have chosen to set things up such that $\sqrt{-g}=e^{2A}$. Let us also introduce matter in the form of a scalar field $\psi(t,r)$ which is isotropic but inhomogeneous. 
When specialised to this space-time and integrating over the isotropic coordinates $\theta$ and $\chi$ (following the principle of symmetric criticality) the action becomes, denoting an $r$-derivative with a prime, and a $t$ derivative with a dot,
\be S_{\mathrm{LTB}} =  \frac{1}{2} \int e^A \left(e^{2 B} \left(\left(A'+B'\right) \left(3 A'+B'\right)-\left(\psi '\right)^2\right)-3 \dot{A}^2+4 e^{-B}+3 \dot{B}^2+\dot{\psi }^2\right) \d r \d t \ee
Note that we have eliminated boundary terms. As with the case in general, we see that the conformal factor, $A$, has a scaling symmetry under $\D=\frac{\partial}{\partial A}$. 
The Euler-Lagrange equations are:
\ba \label{LTBEL} &e&^{2 B} \left(3 A''+6 A' B'+\frac{3 A'^2}{2}+2 B''+\frac{7 B'^2}{2}+\frac{\psi '^2}{2}\right)-3 \ddot{A}-\frac{1}{2} 3 \dot{A}^2-2 e^{-B}-\frac{3 \dot{B}^2}{2}-\frac{\dot{\psi }^2}{2}=0 \nonumber \\ 
&e&^{2 B} \left(2 A''+A' B'-A'^2+B''+B'^2+\psi '^2\right)+3 \dot{A} \dot{B}+3 \ddot{B}+2 e^{-B} =0 \nonumber \\
&e&^{2 B} \left(-A' \psi '-2 B' \psi '-\psi ''\right)+\dot{A} \dot{\psi }+\ddot{\psi } =0  \ea
We can now demonstrate the new Herglotz Lagrangian. Following the reduction process, for this metric $L^H$ is 
\be \label{LTBHL} L^H = -2 e^{2 B} B''-\frac{7}{2} e^{2 B} \left(B'\right)^2-\frac{1}{6} e^{2 B} s_r^2-\frac{1}{2} e^{2 B} \left(\psi '\right)^2+\frac{3 \dot{B}^2}{2}+2 e^{-B}+\frac{s_t^2}{6}+\frac{\dot{\psi }^2}{2} \ee
Here I have explicitly retained a second order term, $B''$, which we would usually replace with a first order term and a boundary term, using
\be -2e^{2B} B'' = \partial_r (-2e^{2B} B') + 4e^{2B} B'^2 \ee
However in the Herglotz Lagrangian setup, the total derivative contributes to the action density, since $\partial_a s^a = L^H$. Therefore not only must we include the second term, proportional to $B'^2$, we must also alter the action density $s^r \rightarrow s^r - 2e^{2B}B'$, hence we arrive at:
\be L^H_{\mathrm{LTB}} = -e^{2B} \left(\frac{B'^2}{6} + \frac{\psi'^2}{2}\right) + \frac{3\dot{B}^2}{2}+  \frac{2B' s^r}{3} +\frac{s^{t2}}{6} + \frac{\dot{\psi}^2}{2} + 2e^{-B} - e^{-2B}\left(\frac{s^{r2}}{6}+\frac{\psi'^2}{2} \right) \ee
and here we note that $\partial_a s^a = L^H$ is equivalent to the Euler-Lagrange equation for $A$, under the identification $s^t = -3 \dot{A}, s^r = e^{2B}(3A'+2B')$. It is a straightforward exercise to demonstrate the remaining Euler-Lagrange equations are given by the Herglotz equations for this Lagrangian together with this identification.
We identify the momenta, and thus see
\be \dot{B} =\frac{P_B^t}{3} \quad B' = e^{-2B}(2s^r-3P_B^r) \quad \dot{\psi}=P_\psi^t \quad \psi'=e^{-2B}P_\psi^r \ee
The contact Hamiltonian for our system is then
\be H^c= e^{-2 B} \left(2 P_B^r s^r-\frac{3 P_B^{r2}}{2}-\frac{P_\psi^{r2}}{2}-\frac{s^{r2}}{2}\right)-2 e^{-B}+\frac{P_B^{t2}}{6}+\frac{P_\psi^{t2}}{2}-\frac{s^{t2}}{6} \ee
From which the contact Hamilton equations for $B$ and $\psi$ recover their Euler-Lagrange counterparts, and the contact equation for the action density reproduces the Herglotz Lagrangian, and hence the Euler-Lagrange equation for $A$.

\section{Discussion}
\label{Sec:Conclusions}

Let us revisit the main results of this article. I have shown that where there exists a scaling symmetry, both Lagrangian and Hamiltonian descriptions of field theories have an equivalent representation in terms of Herglotz Lagrangian and k-contact Hamiltonian field theories respectively.  These representations are independent of some choice of `scale' within the system, which when changed does not effect the observable properties of the system itself. I have applied these results to general relativity wherein I show that changes to the conformal factor which do not change its derivatives (i.e. a global translation of $\phi \rightarrow \phi + \mu$ for some constant $\mu$) can be eliminated from the system. In doing so we recover a Herglotz Lagrangian system with dynamical variables corresponding to the conformally invariant contributions to the metric. In this the dynamics of the derivatives of the conformal factor are reproduced by the action densities. 

In mechanical theories Herglotz Lagrangians and contact Hamiltonians are used to describe systems with friction. This frictional behaviour is manifest through the non-conservation of the Hamiltonian itself, and through the non-conservation of a symplectic two-form on phase space. In the cosmological sector, this can be seen as the `Hubble friction' through which the expansion of the universe changes the dynamics of fields present. The same is true of the conformal factor more generally; if we consider adding a minimally coupled massless scalar field to the Einstein-Hilbert action, equation \ref{EHPhi}, with Lagrangian
\be L_\psi = \frac{1}{2} e^\phi \gamma^{ab} \partial_a \psi \partial_b \psi \ee
Then the equation of motion for $\psi$ is 
\be \partial_a \left( e^\phi \gamma^{ab} \partial_b \psi \right) = 0 \rightarrow  \partial_a \left( \gamma^{ab} \partial_b \psi \right)= - \gamma^{ab} \partial_a \phi \partial_b \psi \ee
where the right hand side would be zero if the conformal factor were constant, and would hence reproduce the dynamics of a free scalar field on flat space. The role of the conformal factor appears akin to friction, altering the conservation law.

We can add the massless scalar to the Herglotz Lagrangian, equation \ref{LHGR}, through the additional term
\be L^H_\psi = \frac{1}{2} \gamma^{ab} \partial_a \psi \partial_b \psi \ee
which corresponds to the additional term that we would arrive at by reducing the scalar field Lagrangian added to the Einstein-Hilbert action. The equation of motion for this field is
\be \partial_a \left( \gamma^{ab} \partial_b \psi \right) = - \frac{1}{3} s^a \partial_a \psi \ee
which reproduces that derived from the Einstein-Hilbert action when we apply the definition of $s^a$. 

As we have seen, the role of the conformal factor is reproduced by the action density, which has a friction-like property. Of course, this is not strictly frictional; the phase space volumes occupied by solutions may increase as well as decrease, as can the Hamiltonian of a field. This does allow for an alternative interpretation of GR when viewed as an operational theory. As the dynamics of observables are entirely equivalent, nothing from this representation contradicts the interpretation of GR as the dynamics of geometry. In the cosmological sector, the change of the conformal factor over time is consistent with an interpretation as the expansion of a spatial slice. However, when viewed through the lens of a frictional system, another entirely physically consistent interpretation is also possible. In this interpretation the conformal factor is simply not present in the first place. As such the dynamics appears consistent with a description in terms of a metric with fixed determinant and friction. Such an interpretation may be of particular interest in the neighborhood of singularities, particularly those characterized by the vanishing of the conformal factor such as the Schwarzschild interior and the big bang in cosmology. As the conformal factor has been excised from the theory, there is no requirement for its dynamics to be well-defined, which can resolve some of the problems with continuing space-times beyond singularities\cite{koslowski2018through,mercati2019through,sloan2019scalar,adamo2023gauge,hoffmann2024continuation,mercati2022traversing}.

There is also the potential for this reinterpretation of GR to be of interest in the realm of quantum gravity. Canonical quantizations of GR, such as those employed in Wheeler-de Witt quantizations and Loop Quantum Gravity, rely on promoting elements of the metric to quantum operators. The excision of the conformal factor at the classical level indicates that this may not be an ideal choice for promotion to a quantum operator in its own right. The quantization of contact systems is still somewhat of an open issue. Odd dimensional phase spaces do not have symplectic structures which can be naturally converted into commutators, and thus the usual route of finding a representation of a Poisson algebra is no longer directly viable. It may therefore be necessary to follow the techniques of quantization of nonconservative (frictional) systems to find appropriate quantizations of GR. 

\bibliographystyle{ieeetr}
\bibliography{DynSimFields}
\appendix 

\section{Direct Demonstration of Equivalence in General Relativity}
\label{Sec:Appendix}

In section \ref{Sec:GR} I presented the Herglotz Lagrangian density that reproduces the dynamics of GR. This was obtained by using the results of section \ref{Sec:Lag}. Here for completeness I will directly calculate the equations of motion for the dynamical variables in each case and show that they are equivalent. Although this is not strictly necessary, as the equivalence is guaranteed by previous results, doing so illustrates the relationship quite directly. Further, we should note that the Ricci scalar depends on the second derivatives of the metric fields $\gamma$. It is normal to eliminate these terms in favour of first order terms and total derivatives - hence boundary contributions to the action. However, here I will retain second derivative terms as they provide insight into how the results above can be extended to higher derivative theories in general. A full accounting of such theories is the subject of future work.

Let us take as our starting point equation \ref{EHPhi}. First, let us vary the action with respect to terms in $\phi$:

\ba \nonumber \delta S &=& \int e^\phi \left( \delta \phi (R + \frac{3}{2}\gamma^{ab} \partial_a \phi \partial_b \phi) +\delta \partial_a \phi (3 \gamma^{ab} \partial_b \phi) + \delta R + \delta \gamma^{ab} \frac{3}{2}  \partial_a \phi \partial_b \phi \right) \,\d^4 x \\
			&=& \int \d^4 x \delta \phi e^\phi \left( R -\frac{3}{2} \gamma^{ab}\partial_a \phi \partial_b \phi - 3 \partial_a (\gamma^{ab}\partial_b \phi) \right) + \delta \gamma^{ab}(...) + \mathrm{Boundary} \ea

Hence we arrive at 
\be 3 \partial_a (\gamma^{ab} \partial_b \phi) =  R(\gamma) - \frac{3}{2} \gamma^{ab} \partial_a \phi \partial_b \phi \ee
and so if we set $s^a = 3\gamma^{ab} \partial_b \phi$, this is equivalent to the role of $s^a$ as action density for the Herglotz Lagrangian of \ref{LHGR}:
\be L^H = \partial_a s^a = R(\gamma) - \frac{1}{6} \gamma_{ab} s^a s^b \ee
To calculate the equations of motion for $\gamma^{ab}$ let us first note that the Euler-Lagrange equations for actions that include second derivatives of a field $\psi$ are
\be D_a D_b \left( \frac{\partial L}{\partial (\partial_a \partial_b \psi)} \right) - D_a \left(\frac{\partial L}{\partial (\partial_a \psi)} \right) + \frac{\partial L}{\partial \psi} = 0 \ee
which in the case of the our Lagrangian density becomes
\ba \nonumber \label{GammaEoM} \partial_a \partial_b \frac{\partial R}{\partial(\partial_a \partial_b \gamma^{ab})} +\partial_a \partial_b \phi  \frac{\partial R}{\partial (\partial_a \gamma^{cd})} +\partial_a \phi \, \partial_b \left( \frac{\partial R}{\partial (\partial_a \partial_b \gamma^{cd})}  \right) + \partial_b \phi \, \partial_a \left( \frac{\partial R}{\partial (\partial_a \partial_b \gamma^{cd})}  \right) \\
 +\partial_a \partial_b \left( \frac{\partial R}{\partial (\partial_a \partial b \gamma^{cd})}  \right) -\partial_a \phi \frac{\partial R}{\partial (\partial_a \gamma^{cd})} - \partial_a \left( \frac{\partial R}{\partial (\partial_a \gamma^{cd})} \right) +  \frac{\partial R}{\partial \gamma^{cd}} + \frac{3}{2} \partial_c \phi \partial_d \phi  = 0 \ea

The Herglotz Lagrange equations for a theory that has second derivatives of a field $\psi$ are equivalent to the Euler-Lagrange equations replacing $D_a$ by $D^L_a$, where
\be D^L_a f = D_a f - \frac{\partial L^H}{\partial s^a} f \ee
Note that $D^L_a$ is not a derivative - it does not obey Leibniz's rule. Following our process for obtaining the Herglotz-Lagrange equations we obtain
\ba \nonumber D^L_a \left( \partial_b  \left(\frac{\partial R}{\partial (\partial_a \partial_b \gamma^{cd})} \right) + \frac{1}{3} \gamma_{be} s^e \frac{\partial R}{\partial (\partial_a \partial_b \gamma^{cd})} \right) - \partial_a \left( \frac{\partial R}{\partial(\partial_a \gamma^{cd})}\right) \\ +\frac{1}{3} \gamma_{ae} s^e \frac{\partial R}{\partial(\partial_a \gamma^{cd})} + \frac{\partial R}{\partial \gamma^{cd}} - \frac{1}{6} s^a s^b \gamma_{ac} \gamma_{ad} = 0  \ea
and further expanding the initial term this becomes
\ba \nonumber \partial_a \partial_b \left(\frac{\partial R}{\partial(\partial_a \partial b \gamma^{cd})} \right) + \partial_a \left(\frac{1}{3} \gamma_{be} s^e \frac{\partial R}{\partial (\partial_a \partial_b \gamma^{cd})}\right)+\frac{1}{3} \gamma_{ae} s^e \partial_b \left( \frac{\partial R}{\partial( \partial_a \partial b \gamma^{cd})} \right) \\ + \frac{1}{9} \gamma_{ae}\gamma_{bf} s^e s^f \frac{\partial R}{\partial(\partial_a \partial_b \gamma^{cd})}  +\frac{1}{3} \gamma_{ae} s^e \frac{\partial R}{\partial(\partial_a \gamma^{cd})} + \frac{\partial R}{\partial \gamma^{cd}} - \frac{1}{6} s^a s^b \gamma_{ac} \gamma_{ad} = 0  \ea
It is then a simple exercise to replace the $s^a$ with $3 \gamma^{ab}\partial_b \phi$ to recover the exact same equation of motion as \ref{GammaEoM}. 

Thus I have shown explicitly in the case of GR that the equations of motion obtained from the Herglotz Lagrangian are exactly the same as those of the Einstein-Hilbert action. 
 
\end{document}